\documentclass[12pt]{article}

\begin{document}

\sloppy
\begin{flushright}{SIT-HEP/TM-14}
\end{flushright}
\vskip 1.5 truecm
\centerline{\large{\bf Topological hybrid inflation in brane world}}
\vskip .75 truecm
\centerline{\bf Tomohiro Matsuda
\footnote{matsuda@sit.ac.jp}}
\vskip .4 truecm
\centerline {\it Laboratory of Physics, Saitama Institute of
 Technology,}
\centerline {\it Fusaiji, Okabe-machi, Saitama 369-0293, 
Japan}
\vskip 1. truecm
\makeatletter
\@addtoreset{equation}{section}
\def\theequation{\thesection.\arabic{equation}}
\makeatother
\vskip 1. truecm

\begin{abstract}
\hspace*{\parindent}
In this paper we show that certain types of cosmological brane
configurations may induce topological inflation.
In generic models for topological inflation, there is a strict bound
for the potential of the inflaton field.
Naively, it seems that these constraints can be removed if
inflation starts with a hybrid potential.
However, it is difficult to combine topological inflation and hybrid
inflation, because of the difference in their initial conditions.
In this paper, we show some examples in which a novel type of topological
inflation is realized with the unconventional hybrid potential.
\end{abstract}

\newpage
\section{Introduction}
\hspace*{\parindent}
The idea of topological inflation is independently pointed out by
Vilenkin\cite{original_vilenkin} and Linde\cite{original_linde}, and
recently applied to the scenario of the brane world\cite{branden}.
Let us begin with a simple potential where a scalar field $\phi$
is a one-component scalar field with a double-well potential,
\begin{equation}
V(\phi)=\frac{1}{4}\lambda\left( \phi^2 - \eta^2 \right)^2.
\end{equation}
This model has two minima at $\phi =\pm \eta$, which are degenerated.
Supposing a random state for $\phi$ with a dispersion larger than
$\eta$, the Universe will split into domains of $\phi=+\eta$ and
$\phi=-\eta$.
The cosmological domain walls appear at the boundary 
interpolating between different vacua.
There is the false vacuum state
of $\phi\simeq 0$ in the core of domain walls.
If the size of the false vacuum region is greater than the horizon size,
this region will undergoes inflational expansion.
The conditions for the above simple potential implies $\eta
>M_p$\cite{original_vilenkin, original_linde}.
Although it seems plausible that a topological inflation takes place when
the above condition is satisfied, the constraints for the potential seem
rather crucial, because it requires fine-tuning of the parameter
$\lambda<10^{-12}$.
Even if one considers a supersymmetric flat potential,
it seems difficult to meet the conditions for successful
inflation.
In generic cases, the negative mass of the flat potential
is determined by the gravitino mass $m_{3/2}$\cite{moduli_defect}.
Then the energy density during inflation is almost fixed at 
$V_0\simeq m_{3/2}^2\eta^2$, if single-field inflation is assumed.
The above constraints may be removed if inflation is realized by the
hybrid potential.
In this paper we show examples for topological inflation combined with
hybrid inflation, in which the hybrid potential is realized in the scenario 
of non-tachyonic brane inflation\cite{matsuda_non-tachyonic}. 
\footnote{In general, inflation with low fundamental scale is very
difficult\cite{low_inflation}. Baryogenesis in models with low
fundamental scale is discussed in ref.\cite{low_baryo}.}

\section{Topological brane inflation}
\hspace*{\parindent}
Here we first consider an effective potential of the form,
\footnote{
Here we consider the case where massless or massive modes are
localized on the branes.
The forms of their wave-functions will be gaussian that decays
exponentially in the transverse directions.
For example, we consider the form;
\begin{equation}
\psi(r)\sim e^{(M_1 r)^2},
\end{equation}
where $r$ denotes the transverse distance from the center of the brane, 
and $M_1^{-1}$ is the width of the
wave-function.
When branes are separated, the couplings between localized modes from
different branes will decay exponentially.
On the other hand, when branes are on top of each other, the couplings 
are restored.
We will discuss some examples in the next section, where the couplings
are needed to maintain supersymmetry in the inflation sector.
(Of course the phenomenological supersymmetry breaking remains.
However, in generic cases, the scale of the phenomenological
supersymmetry breaking is much smaller than the required 
supersymmetry breaking during inflation. ) 
If the couplings are responsible for the supersymmetry on the 
branes, one will find supersymmetry restoration on the branes when
branes are coincident.
Let us consider the case when the constant energy density $\sim M^4$
appears when branes are well separated.
The origin of the vacuum energy is the spontaneous breaking of 
supersymmetry on the branes, which is due to the absence of some 
couplings.
If the energy density disappears (i.e., supersymmetry is restored) 
when the interactions appear,
one will find the potential of the form 
\begin{equation}
V(\phi)=M^4\left(1-e^{-(M_1 r)^2} 
\right),
\end{equation}
where $r$ denotes the distance between branes.
In the next section, we will show some examples where supersymmetry
restoration occurs at two points in the extra dimensions.
}
\begin{equation}
\label{simple}
V(\phi)=M^4\left(1-e^{-(M_1^{-1} (\phi+\phi_0))^2} 
-e^{-(M_1^{-1} (\phi-\phi_0))^2}
\right).
\end{equation}
The potential (\ref{simple}) has two minima at 
$\phi=\pm\phi_0$, which are degenerated.
Assuming $|\phi_0|>M_1$,
the peculiar feature of the potential is the flat direction.
The potential is flat except for the small region around two minima.
Although it remains flat at the tree level, the gravitational effect
will lift the potential.
For example, the gravitational correction may take the form,
\begin{equation}
\label{gra_pot}
\Delta V =-m_{3/2}^2\phi^2 +A_n\frac{\phi^{2n+4}}{M_p^{2n}},
\end{equation}
where the first term is the familiar soft mass term, and the second
represents nonrenormalizable terms.
For simplicity, here we have assumed that the soft mass appears with the
negative sign, and the stability of the potential is ensured by the
second term. 
Obviously, the potential (\ref{gra_pot}) induces another (local or
global) minima for the field $\phi$.
Here we simply assume that the global vacua are at $\phi=\pm \phi_0$,
and denote the additional local minima by $\phi_{local}^i$.
The local minima $\phi_{local}^i$ are assumed to be located outside the
region of $\phi < |\phi_0|$.
Supposing a random initial state for the field $\phi$, the Universe is
split 
into domains of $\phi=0$, $\phi=\phi_0$ and $\phi_{local}^i$.
In this case, the width of the domain walls interpolating
between different vacua is about $\delta \sim m_{3/2}^{-1}$,
which is not related to the vacuum energy in the interior 
of the wall.

The obvious advantage of the above potential is the followings.
Since there is no relation between the vacuum energy in the interior of
the wall and the width of the domain wall,
it is possible to find a model in which the vacuum expectation value of the
inflaton field does not exceed the Planck scale.
At the same time, the requirement from the COBE measurements of the
microwave background anisotropies is crudely
\begin{equation}
\frac{\delta \rho}{\rho}\simeq \frac{H^3}{V'}
\simeq \frac{M^6}{M_p^3 \phi_{COBE}m_{3/2}^2}\simeq 10^{-5}.
\end{equation}
Here $\phi_{COBE}$ denotes the expectation value of $\phi$ when the
COBE scales leave the horizon.
Obviously, the above requirement is much looser than the conventional one.
One may critically think that the fine-tunings of the parameter is 
simply replaced by the fine-tunings of the scales.
In the context of the string theory, the consistency of the model
is ensured by the requirement of additional dimensions.
In models with large extra dimensions, 
the observed Planck mass is obtained by the relation $M_p^2=M^{n+2}_{*}V_n$,
where $M_{*}$ and $V_n$ denote the fundamental scale of gravity
and the volume of the $n$-dimensional compact space.
Assuming more than two extra dimensions, $M_{*}$ may be
close to the TeV scale without conflicting any observable bound.
The most natural embedding of this picture in
the string theory context will be realized by the brane construction.
Thus in the brane world scenario, there is no obvious reason to 
believe that the fundamental scale is as high as the Planck scale.
In this respect, the fundamental scale is the scale that should be 
determined by observations, rather than the input parameter.
We believe that the fundamental scale should be determined by the
cosmological requirements.

Our goal in this paper is to find a concrete example, in which the
above idea is realized within the natural settings of the brane world.

\subsection{Non-tachyonic brane inflation due to the D-term}
\hspace*{\parindent}
Here we consider non-tachyonic brane inflation in
ref.\cite{matsuda_non-tachyonic},
and examine whether one can construct a model that is suitable for our
purposes.
In general, F-term inflation suffers from the old serious difficulty.
In the past, the idea of D-term inflation was invoked to solve the
problem of F-term inflation in conventional supergravity.
Our model for inflation is a modification of the conventional
hybrid inflation.
Here we consider an extra dimension $(y)$ and a localized Fayet-Iliopoulos
term placed at $y=0$ of the form
\begin{equation}
\label{FI}
\xi D \delta(y),
\end{equation}
where $D$ is an auxiliary field of the vector superfield.
We consider an additional abelian gauge group $U(1)_X$ in the bulk,
while the Fayet-Iliopoulos term for $U(1)_X$ is localized on a brane. 
We also include the fields $\phi_X$ that has the $U(1)_X$ charge and
localized on the other brane at $y=y_1$.
When two branes are located at a distance, $|y_1| >>M_*^{-1}$,
the Fayet-Iliopoulos term (\ref{FI}) breaks supersymmetry on the 
brane.\footnote{Here we temporally ignore the derivative terms. }
In this case, as in the conventional models for brane inflation, 
the inflaton field is the moduli that parametrizes the
brane distance.
Here we denote the moduli by $\phi=M_*^2 y_1$,
where $M_*$ is the fundamental scale.
The potential for the inflaton field is flat at the tree level.
As we are considering D-term inflation, the mass of the inflaton
($m_{\phi}$) may be much smaller than the Hubble parameter.
Then a modest limit is $m_{\phi} \ge m_{3/2}$,
where  $m_{3/2}$ is the gravitino mass in the true vacuum.
If the extra dimension is $S^1$, the resultant domain wall is the $N=1$
axionic domain wall that is bounded by the strings.
The potential is periodic, and similar to
eq.(\ref{simple}). 

\underline{model 2}

Let us consider the next example, five-dimensional theory that is
made chiral by choosing the right boundary conditions\cite{orbifold5d}.
An abelian gauge multiplet of the five-dimensional gauge sector consists
of a vector superfield $V$ whose components are the four-dimensional
part of the vector gauge field $A^{\mu}$, the left-handed gaugino,
auxiliary field $D$, and a chiral scalar field $\Phi$.
The lowest component of $\Phi$ is a complex scalar
$\phi=(\Sigma+iA_5)/\sqrt{2}$, where $A_5$ is the fifth component of the
vector field.
The five-dimensional Lagrangian density is given by
\begin{equation}
\left[\frac{1}{g^2}\left(\Phi^{\dagger}\Phi
-\sqrt{2}(\Phi^{\dagger}\Phi)\partial_y V -V\partial_y^2 V
\right)\right]_{\theta^4}
+\left[W_\alpha W^\alpha \right]_{\theta^2}+h.c.
\end{equation}
We assume a Fayet-Iliopoulos term on a brane at $y=0$, which looks like
\begin{equation}
\left[2\xi V \delta(y)\right]_{\theta^4}.
\end{equation}
Here we assume that the fields $\phi_X^{\pm}$, which have charges of
$\pm q_X$, are  localized  on a  
brane at $y=L/2$. 
The D-flat condition is
\begin{equation}
-D=\left[2 \xi \delta(y) +\frac{g q_X}{2} (|\phi_X^{+}|^2 -
|\phi_X^{-}|^2)\delta(y-L/2) + \partial_y \Sigma \right]=0
\end{equation}
which is satisfied even though the fields  $\phi_X^{\pm}$ are not 
located at the brane on which the Fayet-Iliopoulos term is localized.
In this case, the explicit form of the solution is 
$|\phi_X^{+}|^2 -|\phi_X^{-}|^2=-4\xi/g q_X$, $\Sigma=\xi
\epsilon(y)$.
This simple example shows that there is a possibility that supersymmetry 
can be restored by the derivative terms even if the 
Fayet-Iliopoulos term and the charged matter field are separated.
Such a configuration is possible for five-dimensional models with
orbifolded boundary conditions, at least when the Fayet-Iliopoulos term
and the charged matter are located exactly at different fixed points.
Then we can find two degenerated supersymmetric vacua at each fixed point. 
The potential for the moduli field $\phi$ is
\begin{equation}
V(\phi)=M^4\left(1-e^{-\left(\frac{\phi}{M_*} \right)^2} 
-e^{-\left(\frac{\phi-M_*^2 L/2}{M_*}\right)^2}\right).
\end{equation}
Here the field $\phi$ denotes the moduli for the location of the
moving brane on which the fields $\phi_X^{\pm}$ are localized.
The above potential has two degenerated minima at $\phi=0$ and 
$\phi=M_*^2 L/2$.
The potential between the degenerated vacua is almost flat at the tree
level, however it may be lifted by the conventional supersymmetry
breaking.
The situation is precisely the same as eq.(\ref{simple}).

\section{Conclusions and Discussions}
\hspace*{\parindent}
In this paper we have considered cosmological brane
configurations that induce topological inflation.
In generic models for topological inflation, there is a strict bound.
In this paper, we show some examples in which a novel type of topological
inflation is realized with the unconventional hybrid potential, which
avoids the difficulties of the conventional topological inflation.

\section{Acknowledgment}
We wish to thank K.Shima for encouragement, and our colleagues in
Tokyo University for their kind hospitality.


\begin{thebibliography}{1}
\bibitem{original_vilenkin}
A. Vilenkin, Phys.Rev.Lett.72:3137-3140,1994. 
\bibitem{original_linde}
A. D. Linde, Phys.Lett.B327:208-213,1994.
\bibitem{branden}
S. Alexander, R. Brandenberger, M. Rozali,
``Nontopological inflation from embedded defects'', hep-th/0302160 
\bibitem{moduli_defect}
K. Freese, T. Gherghetta, H. Umeda, Phys.Rev.D54:6083-6087,1996.
\bibitem{low_inflation}
N. Arkani-Hamed, S. Dimopoulos, N. Kaloper, and J. March-Russell,
Nucl.Phys.B567:189-228,2000;
R. N. Mohapatra, A. Perez-Lorenzana, and C. A. de S. Pires,
Phys.Rev.D62:105030,2000;
A. M. Green and A. Mazumdar, Phys.Rev.D65:105022,2002;
T. Matsuda, Phys.Rev.D66:107301,2002; Phys.Lett. B486 (2000) 300-305;
D. H. Lyth, Phys.Lett.B448:191-194,1999; Phys.Lett.B466:85-94,1999.
\bibitem{low_baryo}
G.R. Dvali, G. Gabadadze, Phys.Lett.B460:47-57,1999;
T. Matsuda, Phys.Rev.D65:103502,2002,
Phys.Rev.D66:023508,2002, Phys.Rev.D65:107302,2002,
Phys.Rev.D66:047301,2002; Phys.Rev.D64:083512,2001;
J.Phys.G27:L103-L108,2001; 
A. Masiero, M. Peloso, L. Sorbo, and R. Tabbash, Phys.Rev.D62:063515,2000;
A.Mazumdar, Nucl.Phys.B597(2001)561, Phys.Rev.D64(2001)027304;
A. Mazumdar and A. Perez-Lorenzana, Phys.Rev.D65:107301,2002; 
R. Allahverdi, K. Enqvist, A. Mazumdar, and A. Perez-Lorenzana,
Nucl.Phys.B618:277-300,2001;
A.Pilaftsis, Phys.Rev.D60:105023,1999;
R.Allahverdi, K.Enqvist, A.Mazumdar and A.P-Lorenzana,
Nucl.Phys. B618:377,2001;
S. Davidson, M. Losada, and A. Riotto, Phys.Rev.Lett.84:4284-4287,2000.
\bibitem{matsuda_non-tachyonic}
T. Matsuda, ``Non-tachyonic brane inflation'', hep-ph/0302035;
``F-term, D-term and hybrid brane inflation'', hep-ph/0302078.
``Thermal hybrid inflation in brane world'', hep-ph/0302253;
Phys.Rev.D65:103501,2002; Phys.Lett. B423 (1998) 35-39.
\bibitem{orbifold5d}
N. Arkani-Hamed, T. Gregoire, J Wacker, JHEP 0203:055,2002 
H-C. Cheng, B. A. Dobrescu, C. T. Hill, Nucl.Phys.B589:249-268,2000 
D. E. Kaplan, T. M.P. Tait, JHEP 0111:051,2001 
\end{thebibliography}
\end{document}